\documentclass[aps,prx,twocolumn,amsmath,amssymb,nofootinbib,superscriptaddress,floatfix,epsfig]{revtex4-1}
\usepackage{graphicx}
\usepackage{tabularx}
\usepackage{subfigure}
\usepackage{epsfig}
\usepackage{dcolumn}
\usepackage{bm}
\usepackage{euscript}
\usepackage{ulem}
\usepackage{color}
\usepackage{epsfig,psfrag,subfigure}
\usepackage{amsfonts}
\usepackage{exscale}
\usepackage{amsbsy}
\usepackage{ulem}
\usepackage{setspace}

\usepackage{graphicx}
\usepackage{dcolumn}
\usepackage{bm}

\usepackage{braket}
\usepackage{mathrsfs}

\begin{document}
\title{Visualizing Molecular Orientational Ordering and Electronic Structure \\in Cs$_n$C$_{60}$ Fulleride Films}
\author{Sha Han}
\affiliation{State Key Laboratory of Low-Dimensional Quantum Physics, Department of Physics, Tsinghua University, Beijing 100084, China}
\author{Meng-Xue Guan}
\affiliation{Institute of Physics, Chinese Academy of Sciences, Beijing 100190, China}
\author{Can-Li Song}
\thanks{Corresponding author: clsong07@mail.tsinghua.edu.cn}
\affiliation{State Key Laboratory of Low-Dimensional Quantum Physics, Department of Physics, Tsinghua University, Beijing 100084, China}
\affiliation{Frontier Science Center for Quantum Information, Beijing 100084, China}
\author{Yi-Lin Wang}
\affiliation{Center of Nanoelectronics and School of Microelectronics, Shandong University, Jinan 250100, China}
\author{Ming-Qiang Ren}
\affiliation{State Key Laboratory of Low-Dimensional Quantum Physics, Department of Physics, Tsinghua University, Beijing 100084, China}
\author{Sheng Meng}
\affiliation{Institute of Physics, Chinese Academy of Sciences, Beijing 100190, China}
\author{Xu-Cun Ma}
\thanks{Corresponding author: xucunma@mail.tsinghua.edu.cn}
\affiliation{State Key Laboratory of Low-Dimensional Quantum Physics, Department of Physics, Tsinghua University, Beijing 100084, China}
\affiliation{Frontier Science Center for Quantum Information, Beijing 100084, China}
\author{Qi-Kun Xue}
\affiliation{State Key Laboratory of Low-Dimensional Quantum Physics, Department of Physics, Tsinghua University, Beijing 100084, China}
\affiliation{Frontier Science Center for Quantum Information, Beijing 100084, China}
\affiliation{Beijing Academy of Quantum Information Sciences, Beijing 100193, China}

\date{\today}
\begin{abstract}
Alkali-doped fullerides exhibit a wealth of unusual phases that remain controversial by nature. Here we report a cryogenic scanning tunneling microscopy study of the sub-molecular structural and electronic properties of expanded fullerene C$_{60}^{n-}$ films with various cesium (Cs) doping. By varying the discrete charge states and film thicknesses, we reveal a large tunability of orientational ordering of C$_{60}^{n-}$ anions, yet the tunneling conductance spectra are all robustly characteristic of energy gaps, hallmarks of Jahn-Teller instability and electronic correlations. The Fermi level lies halfway within the insulating gap for stoichiometric Cs doping level of $n$ $=$ 1, 2, 3 and 4, apart from which it moves toward band edges with concomitant electronic states within the energy gap. Our findings establish the universality of Jahn-Teller instability, and clarify the relationship among the doping, structural and electronic structures in Cs$_n\textrm{C}_{60}$ fullerides.
\end{abstract}
\maketitle

\section{Introduction}
Emergent phenomena and exotic electronic phases in organic molecular solids, including superconductivity \cite{hebard1991potassium,mitsuhashi2010superconductivity,ardavan2011recent,kubozono2016recent}, magnetism \cite{allemand1991organic,chen2016symmetry}, charge \cite{charge2000chow} and orientational ordering \cite{Orientational1991Heiney,Novel2007Wang,Rossel2011Growth,tang2012orientational,jung2014atomically}, have been routinely explored in condensed matter physics, as well as for potential applications in electronics \cite{dimitrakopoulos2002organic}. Alkali-doped fullerides ($A_3$C$_{60}$), for example, exhibit superconductivity with a maximum transition temperature $T_\textrm{c}$ $\sim$ 40 K and bring about special interest in the superconductivity research community \cite{ganin2008bulk,takabayashi2009disorder,ganin2010polymorphism,Wzietek2014NMR,zadik2015optimized,nomura2015unified}, because they exhibit a proximity magnetic Mott-insulating state accompanied with a dome-shaped dependence of superconductivity upon pressure or chemical doping in common with unconventional high-$T_\textrm{c}$ superconductors of the cuprate and ferropnictide classes \cite{takabayashi2016unconventional, keimer2015quantum}. A minor but significant distinction is the unique occurrence of on-molecule electron-vibration coupling in fullerides, which might lift the active $t_{1u}$ orbital degeneracy of spherical C$_{60}^{n-}$ anions via Jahn-Teller (JT) distortions and competes with local exchange coupling \cite{klupp2012dynamic,potovcnik2014jahn}, thereby leading to an orbital disproportionation of filled electrons \cite{Intramolecular1997Ceulemans, iwahara2016orbital,Hoshino2017spontaneous,Ishigaki2018Spontaneously,Isidori2019charge}. The complexity associated with comparable bandwidth to the energy scale of on-molecule electron-phonon coupling and electronic correlations (characterized by the Hubbard $U$) renders it nontrivial to perceive comprehensively the fundamental electronic structure and microscopic mechanism of fulleride superconductivity on the verge of Mott localization \cite{yang2003band}.

On the other hand, the C$_{60}$ molecules are orientationally disordered in the face-centered cubic trivalent fulleride superconductors $A_3$C$_{60}$ ($A$ = K, Rb and Cs) \cite{hebard1991potassium, kubozono2016recent,ganin2008bulk,takabayashi2009disorder,ganin2010polymorphism,Wzietek2014NMR,zadik2015optimized}. These merohedral disorders could affect the electronic structure and properties as well that appear unique to the molecular solids \cite{Orientational1992gelfand, Intramolecular1997Ceulemans,Orientational1992gelfand}. Furthermore, the molecular orientation depends substantially on charge state of the C$_{60}^{n-}$ anions \cite{wachowiak2005visualization,wang2008tuning,daughton2011orientation,bozhko2013correlation}. As thus, it becomes increasingly significant to determine experimentally the doping dependence of molecular configuration and fundamental electronic structure that are essential prerequisites for understanding the novel properties in fullerene-based solids.

Cesium (Cs) doped fullerides are particularly fascinating, because the most expanded crystal structure leads to a relatively small intermolecular hopping. This serves as a model system to understand the interplay between the competing intramolecular interactions and therefore ground states in the orbitally degenerate fulleride families. However, the sample imperfections and phase separation in this category of materials have largely frustrated a clear-cut relationship between the composition stoichiometry and electronic structure \cite{ganin2008bulk,takabayashi2009disorder,ganin2010polymorphism,Wzietek2014NMR,zadik2015optimized}. Local probes may circumvent this issue, but thus far have been limited to a few of under-doped K$_n$C$_{60}$ thin films ($n$ is the doping concentration) on metal substrates \cite{Novel2007Wang,wachowiak2005visualization,wang2008tuning}, where the substrate screening effects play an important role in the electronic structure of fullerides. Here we actualize a systematic doping control of the expanded Cs$_n$C$_{60}$ fulleride films on graphitized SiC(0001) substrates that allows us to track the doping-dependent sub-molecular configuration and fundamental electronic structure by means of scanning tunneling microscopy (STM). Despite a diversity of molecular orientational ordering, the electronic ground states are all insulating in Cs$_n$C$_{60}$ fullerides with integer $n$ $=$ 1, 2, 3 and 4, which we ascribe to the cooperative interplay between the on-molecule Jahn-Teller distortions and electronic correlations.

\section{Experimental methods}
All experiments were carried out in an ultrahigh vacuum low-temperature STM system equipped with a molecular beam epitaxy (MBE) chamber for \textit{in-situ} sample preparation. The base pressure of both MBE and STM chambers is better than 1.0 $\times$ 10$^{-10}$ Torr. Monolayer and multilayer fullerene epitaxial thin films were prepared in the MBE chamber by thermal evaporation of C$_{60}$ source out of standard Knudsen cell on graphitized SiC(0001) substrates with a resistivity of $\sim$ 0.1 $\Omega \cdot$cm. Evaporation of desired equivalents of Cs onto the pristine C$_{60}$ films from a thoroughly outgassed SAES getter at room temperature followed by moderate annealing at $\sim$ 473 K yields electron-doped Cs$_n$C$_{60}$ films with well-controlled doping level $n$. Here the flux rates of C$_{60}$ and Cs sources are estimated by separately depositing them onto a clean graphene substrate at a cryogenic temperature of approximately 100 K and counting the number of C$_{60}$ molecules or Cs from STM images. The calculated ratio of flux and growth duration for Cs and C$_{60}$ are used to determine the nominal Cs doping in Cs-doped fullerene films.

After MBE growth, the fulleride films were $in$-$situ$ transferred into STM head for topographic imaging and electronic structure measurements at 4.8 K. Polycrystalline PtIr tips were pre-cleaned by $e$-beam heating, calibrated on MBE-grown Ag films and used throughout the experiments. All STM topographies were acquired in a constant-current mode with the bias voltage $V$ applied to the sample. Tunneling spectra were measured using the standard lock-in technique with a small bias modulation of 20 meV at 987.5 Hz.

\section{Results and discussion}
\subsection{STM characterization of undoped C$_{60}$}
As substrate, the graphene has its unique superiority in that it decouples effectively C$_{60}$ overlayer from the underlying substrates and enables to deduce readily the innate character of alkali fullerides \cite{jung2014atomically,cho2012structural}. Figure 1(a) depicts a large-scale STM topography of pristine (undoped) C$_{60}$ films on graphene, assembled into a (111)-orientated face-centered cubic crystal structure. We observe the coexistence of monolayer, bilayer, multilayer C$_{60}$ films as well as the graphene/SiC surface (see also inset). Regardless of layer indexing, the fullerenes are closely packed into hexagonal structure with an intermolecular spacing $d$ of 10.0 $\pm$ 0.1 $\textrm{\AA}$ [Figs.\ 1(b) and 1(c)], consistent with previous reports \cite{Rossel2011Growth,tang2012orientational,jung2014atomically,wachowiak2005visualization,wang2008tuning,cho2012structural}. The sub-molecular structure, however, reveals complex configurations of C$_{60}$ molecules with respect to the surface. As typical for fullerene in a spherical basis, five distinct orientations, with a hexagon (H), a pentagon (P), a 6:6 bond (H:H), a 6:5 bond (H:P) and a carbon apex (CA) pointing up, have been routinely observed in experiments \cite{Rossel2011Growth,tang2012orientational,jung2014atomically,Orientational2001wang} and are drawn schematically in Fig.\ 1(d). Here the H:H bonds between two hexagons exhibit a double-bond character, while the H:P bonds separating pentagons and hexagons have a prevalent single-bond character. Given that the 6:5 bonds have relatively lower electron density and the C$_{60}$ pentagons brighten in the empty-state STM topography, one can readily distinguish the orientations of C$_{60}$ by symmetry, and typically superimpose them onto the corresponding C$_{60}$ molecules in Figs.\ 1(b) and 1(c). Apparently, the C$_{60}$ molecules with tri-star-like and two-lob intramolecular structures in Fig.\ 1(c) correspond to the H and H:H orientations, namely with the hexagon and 6:6 bond facing the surface, respectively, whereas for the third kind of C$_{60}$ orientation the brightest spots are always found to be off-centered from the C$_{60}$ center (see the coaxial circles and dots). We therefore assign reasonably them as H:P orientation, and the bright spots are characteristic of the pentagons. In Fig.\ 1(b), except for the H:P orientated C$_{60}$, the other dim molecules are then assigned as P-orientated C$_{60}$. This is understandable since a C$_{60}$ molecule with its bond facing up is higher than that with the pentagon ring facing up. Energetically, considering that the electron-doped graphene will preferentially transfer electrons to undoped C$_{60}$ monolayer, the P orientation is more favorable as well.

\begin{figure*}[t]
    \includegraphics[width=2\columnwidth]{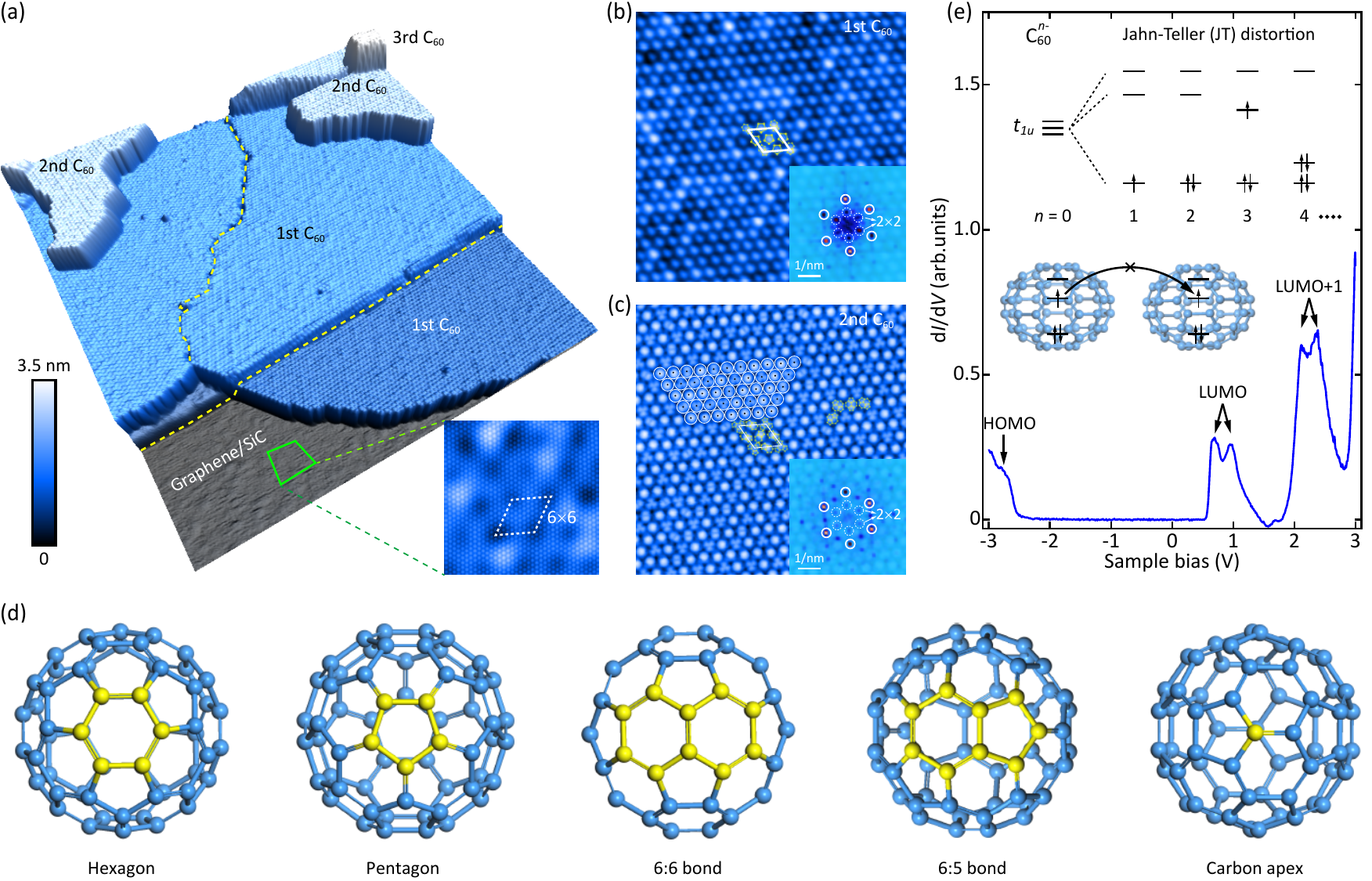}
\caption{Characterization of pristine C$_{60}$ fullerenes on graphene. (a) STM topography (100 nm $\times$ 100 nm) of as-grown C$_{60}$ thin films taken with sample bias $V$ $=$ 3.0 V and tunneling current $I$ $=$ 200 pA. Color-coded plateaus reveal four distinct regions: the C$_{60}$-free graphene/SiC surface (gray), the first layer (blue), second layer (light cyan) and third layer (white) of C$_{60}$ fullerenes. The yellow dashes designate two steps from the substrate. Inset: zoom-in STM topography (8 nm $\times$ 8 nm, $V$ $=$ 50 mV, $I$ $=$ 100 pA) of bilayer graphene on SiC showing the well-known 6 $\times$ 6 superstructure (dashed rhombus). (b),(c) High-resolution STM images (20 nm $\times$ 20 nm) of monolayer and bilayer C$_{60}$ films. Insets display the corresponding FFT patterns. Bright spots circled by the solid and dashed lines stem from the self-assembled C$_{60}$ lattice and quasi 2 $\times$ 2 superstructure in the molecular orientation, respectively. The white rhombus marks the surface unit cell. According to the overlaid lattice mesh (coaxial circles and dots) in (c), the brightest C$_{60}$ molecules are H:P-orientated with the pronounced DOS (i.e.\ pentagon) deviating from the centers of fullerenes. Tunneling conditions are (b) $V$ $=$ 1.0 V, $I$ $=$ 100 pA and (c) $V$ $=$ 2.0 V, $I$ $=$ 100 pA. (d) Ball and stick diagrams of C$_{60}$ in the hexagon, pentagon, 6:6 bond, 6:5 bond and carbon apex orientations, from left to right. (e) Representative conductance $dI/dV$ spectrum of C$_{60}$ films on graphene, with setpoint of $V$ $=$ 1.0 V and $I$ $=$ 100 pA. The splitting of triply degenerate $t_{1u}$ LUMO and $t_{1g}$ LUMO+1 levels is due to the JT distortion of icosahedral symmetric C$_{60}$ molecules upon electron doping of fullerene. Inset: modified molecular orbitals (top panel) of C$_{60}^{n-}$ anions by JT distortion (bottom panel), with its magnitude exaggerated in the molecule model. The up and down arrows denote the up- and down-spin electrons, respectively.}
\end{figure*}

A careful scrutiny of sub-molecule resolution STM images reveals a certain regularity in the molecular orientations. For example, the H-oriented C$_{60}$ molecules in the second layer arrange well into a quasi 2 $\times$ 2 superstructure relative to the intermolecular spacing, marked by the white rhombus in Fig.\ 1(c). The reason why we term it a quasi 2 $\times$ 2 superstructure is that the other C$_{60}$ molecules exhibit no long-range order in view of the molecular orientations. Likewise, local 2 $\times$ 2 superstructure in the H:P orientation, albeit faint, develops on C$_{60}$ monolayer [Fig.\ 1b]. The ordering of molecular orientation is further evidenced by fast Fourier transform (FFT) patterns inserted in Figs.\ 1(b) and 1(c), which, in addition to lattice Bragg spots, exhibit extra spots from the quasi 2 $\times$ 2 superstructure of the molecular orientation, circled by the dashed lines.

Figure 1(e) depicts the typical differential conductance $dI/dV$ spectrum measured on the monolayer C$_{60}$ films. The peaks at -2.7 eV, 0.8 eV and 2.3 eV correspond to the highest occupied molecular orbital (HOMO), the lowest unoccupied molecular orbital (LUMO), and the LUMO$+$1 states of C$_{60}$ fullerene, respectively. A larger HOMO-LUMO gap of 3.5 eV reconciles fairly well with negligible electrostatic screening from the underlying graphene substrate \cite{jung2014atomically,cho2012structural}, in marked contrast to fullerenes grown on metal surface \cite{Charge2004Lu,schull2008spatially}. Yet, it is worth noting that the triply degenerate $t_{1u}$ LUMO and $t_{1g}$ LUMO$+$1 levels split individually into two discrete peaks, hallmark of JT distortion in charged fullerenes (inset of Fig.\ 1(e)) \cite{Auerbach1994electron,Manini1994electron}. Here the charging of C$_{60}$ is prompted by electron transfer from graphene, owing to a large electron affinity of fullerene \cite{huang2014high}. This appears to be a good agreement with the tunneling spectrum where the Fermi level ($E_F$) lies closer to the LUMO [Fig.\ 1(e)].

\subsection{Orientational ordering in Cs$_n$C$_{60}$}
\begin{figure*}[t]
    \includegraphics[width=1.6\columnwidth]{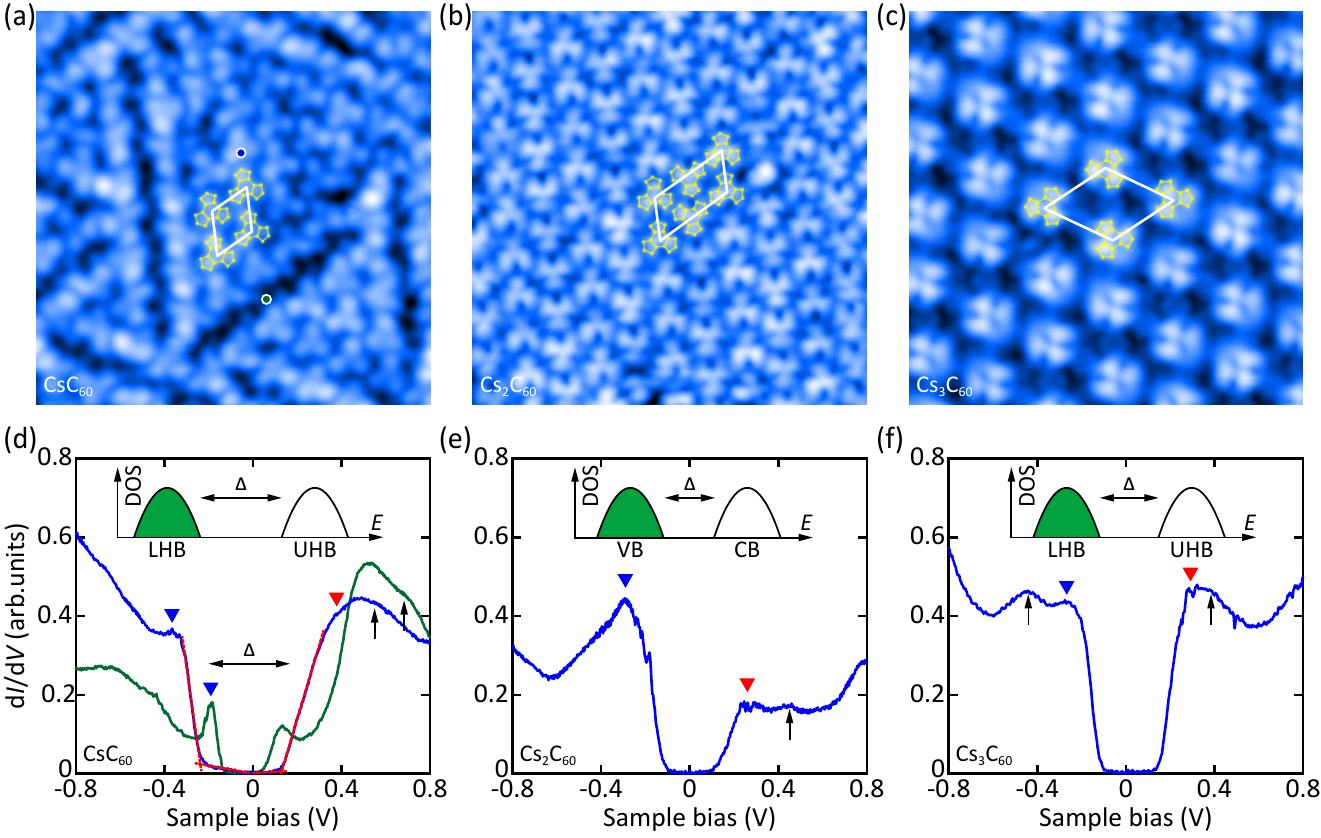}
\caption{Orientational ordering and electronic structure in monolayer Cs$_n$C$_{60}$ for $n$ = 1 $-$ 3. (a)-(c) STM topographic images (10 nm $\times$ 10 nm) and (d)-(f) tunneling $dI/dV$ spectra (setpoint: $V$ $=$ 0.5 V, $I$ $=$ 100 pA) of monolayer C$_{60}^{n-}$ anions with various Cs doping as indicated. Tunneling conditions are (a) $V$ $=$ 1.0 V, $I$ $=$ 100 pA, (b) $V$ $=$ -1.0 V, $I$ $=$ 100 pA and (c) $V$ $=$ 0.65 V, $I$ $=$ 100 pA. The two different spectra in (d) are acquired at the color-coded positions (dots) in (a). Note that the H-orientated fullerenes form 1 $\times$ 2 superstructure and 3 $\times$ 3 superstructure in monolayer Cs$_2$C$_{60}$ and Cs$_3$C$_{60}$, respectively. Inserted in (d)-(f) are schematic band structure of doped fullerenes with only the LHB/VB (green) and UHB/CB (unfilled) shown, matching nicely with the measured conductance spectra. Onsets of occupied (UHB/CB) and unoccupied (LHB/VB) bands are extracted from the intersections of linear fits to the spectral weights just above and below the corresponding band edges, as exemplified by the red dashes in (d). The up arrows denote satellite peaks from JT distortions. The same convention is used throughout.}
\end{figure*}

\begin{figure*}[t]
    \includegraphics[width=1.98\columnwidth]{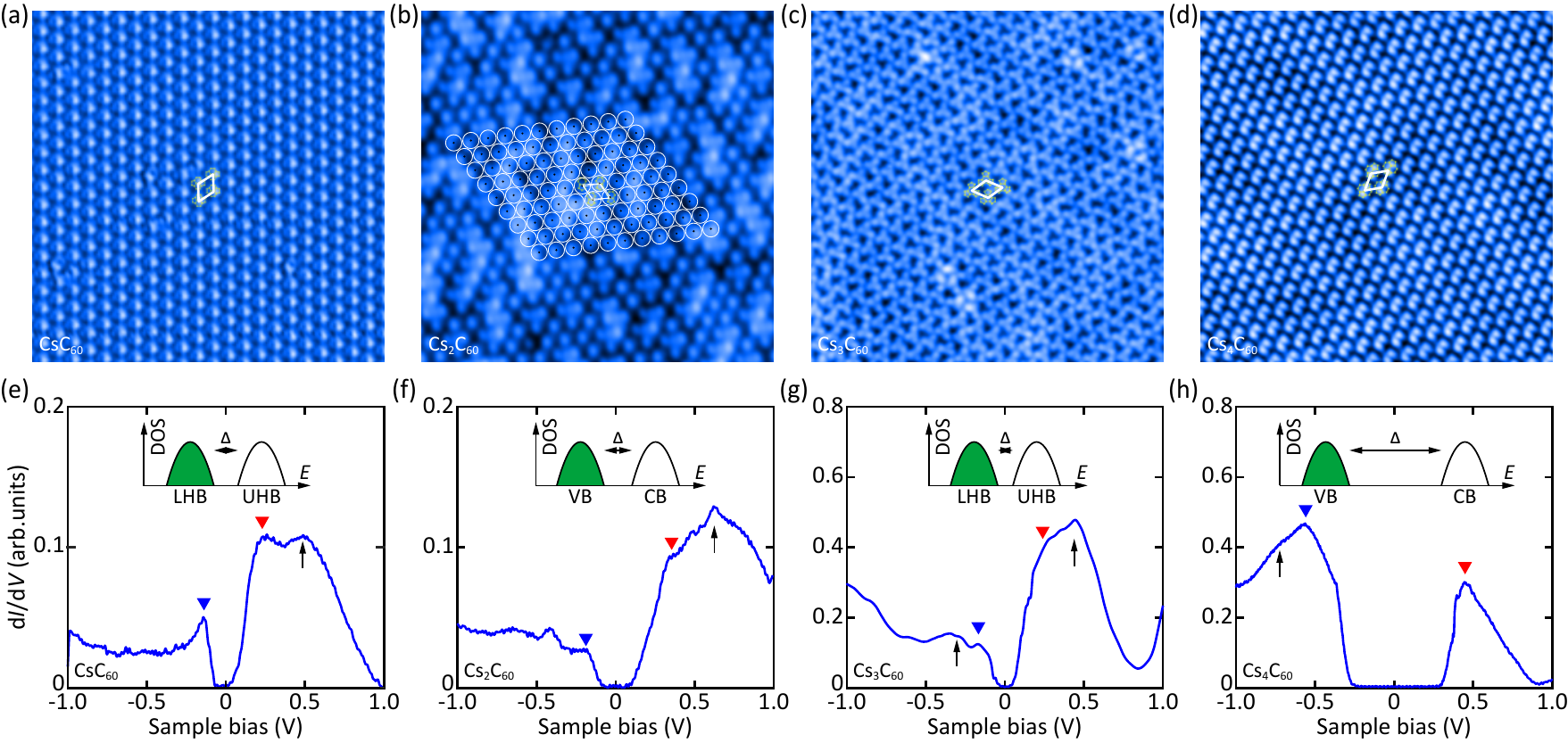}
\caption{Orientational ordering and electronic structure in bilayer Cs$_n$C$_{60}$ for $n$ = 1 $-$ 4. (a)-(d) STM topographic images (20 nm $\times$ 20 nm) and (e)-(h) tunneling $dI/dV$ spectra of bilayer C$_{60}^{n-}$  anions with various Cs doping as indicated. Setpoint: $V$ $=$ 1.0 V, $I$ $=$ 100 pA. The molecular orientational ordering is robustly found, but the configuration of C$_{60}^{n-}$ anions shows dramatic dependence on the charge state $n-$, adopting CA, H:P, H, H:H orientations from $n=1$ to $n=4$. Analogous to undoped C$_{60}$ bilayer, the brightest pentagons in (b) are off-centered and do not conform to a complete hexagonal lattice, for which we assign them to H:P-orientated ordering of C$_{60}$. }
\end{figure*}

We highlight that the molecular orientational ordering is closely associated with charge state of fullerenes (see Appendix A), as convincingly rationalized by a systematic survey of Cs$_n$C$_{60}$ films on the sub-molecular scale. Figures 2 and 3 show STM topographies and electronic structures of Cs$_n$C$_{60}$ films ($n$ = 1, 2, 3 and 4) as functions of the film thickness and electron doping $n-$. Upon further doping, the excess Cs atoms cluster together on the overlayer [Fig.\ 4(a)]. A comparison from STM topographies among STM topographies reveals that the configuration of C$_{60}^{n-}$ anions is hypersensitive to the charge state $n-$ and film thickness. First, a noteworthy observation is the two hook-like intramolecular structure with different strength in bilayer CsC$_{60}$ [Fig.\ 3(a)], which matches with the CA-oriented fullerenes in Fig.\ 1(d). As thus, all the five configurations of C$_{60}$ molecules have been convincingly identified in our study. More remarkably, we reveal that the $n-$ charged fullerenes ($n$ = 1 $-$ 3) are consistently H-orientated in monolayer Cs$_n$C$_{60}$ [Figs.\ 2(a)-(c)], whereas the molecular configurations changes from CA, H:P, H to H:H with increasing $n$ = 1 $-$ 4 in bilayer Cs$_n$C$_{60}$ [Figs.\ 3(a)-3(d)]. Nevertheless, the C$_{60}^{n-}$ anions all exhibit intriguing molecular orientational ordering, despite it being short-ranged in monolayer CsC$_{60}$ [Fig.\ 2(a)]. A closer inspection of the submolecular C$_{60}$ structure reveals the presence of merohedral disorders in stoichiometric Cs$_n$C$_{60}$ fullerides except for bilayer Cs$_4$C$_{60}$. Evidently, the merohedrally disordered molecules arrange into 1 $\times$ 2 and 3 $\times$ 3 superstructures in monolayer Cs$_2$C$_{60}$ [Fig.\ 2(b)] and Cs$_3$C$_{60}$ [Fig.\ 2(c)], whereas no superstructure associated with the merohedral disorders exists in monolayer CsC$_{60}$ [Fig.\ 2(a)] and bilayer Cs$_n$C$_{60}$ ($n$ = 1 $-$ 4) [Figs.\ 3(a)-3(c)].

Apart from stoichiometry with non-integer ratio between Cs and C$_{60}$ constituents, the orientations of C$_{60}$ anions become spatially disordered and the molecular ordering vanishes [Fig.\ 4(b)]. We thus arrive at our first major finding: the molecular orientations of C$_{60}^{n-}$ anions become ordered solely for stoichiometric Cs$_n$C$_{60}$ compounds with integer $n$, including zero. Such finding accords quite nicely with the local orientational ordering in K$_3$C$_{60}$ \cite{wachowiak2005visualization,wang2008tuning}, as well as the preference of molecular orientational ordering in pristine C$_{60}$ multilayers \cite{Rossel2011Growth,tang2012orientational,Orientational2001wang,grosse2016nanoscale} or thin films supported by chemically inert substrates \cite{jung2014atomically,santos2017rotational}. In the latter situations, the C$_{60}$ assemblies on question are only limitedly affected by the substrates and conserve their charge neutrality.

We suggest that the intermolecular Coulomb repulsive interactions between charged C$_{60}^{n-}$ anions play a key role to stabilize the long-range ordering of molecular orientational ordering observed here. The missing molecular ordering in nonstoichiometric Cs$_n$C$_{60}$ [Fig.\ 4(b)] is most likely because the charge states of each fullerene are different and the repulsive forces between them become spatially non-uniform. Further theoretical simulation is desired to thoroughly understand the observed molecular orientational ordering of C$_{60}^{n-}$ anions with varying film thickness and charge state \cite{leaf2016combined}.

\subsection{Absence of metallicity in Cs$_n$C$_{60}$}
Having established the diversities of molecular orientations and their ordering in Cs$_n$C$_{60}$, we below explore the charge state and layer-dependent electronic density of states (DOS) around $E_\textrm{F}$ in Figs.\ 2(d)-(f) (monolayer Cs$_n$C$_{60}$, $n =1-3$) and Figs.\ 3(e)-(h) (bilayer Cs$_n$C$_{60}$, $n =1-4$). Despite somewhat distinction between H-orientated C$_{60}$ domains and the boundaries separating them in monolayer CsC$_{60}$ [Fig.\ 2(d)], the spatial-resolved differential conductance $dI/dV$ spectra are homogeneous in Cs$_n$C$_{60}$ fullerides with the long-range orientational orders. Notwithstanding, all the tunneling spectra present apparent energy gap in the vicinity of $E_\textrm{F}$, consistent with insulating electronic states for Cs$_n$C$_{60}$ fullerides with integer $n$. This constitutes our second major finding. It is also worthy to note that the gaps exist robustly in the off-stoichiometric Cs$_n$C$_{60}$ fullerides [Figs.\ 4(c) and 4(d)]. In order to extract the energy gap, we have reasonably determined onsets of the lowest occupied (marked by red triangles) and the highest unoccupied bands (blue triangles) as the intersections of linear fits to spectral DOS just above and below every band (red dashes) in all $dI/dV$ spectra.

In Fig.\ 5(a), we plot the onset energies (top panel) and extracted band gap $\Delta$ (middle panel) between them versus $n$ and film thickness of Cs$_n$C$_{60}$. Evidently, $\Delta$ not only varies with the charge state $n-$, but also relies critically on film thickness. Despite such alteration, the $E_\textrm{F}$ lies closely at the midgap energy $E_\textrm{i}$ in Cs$_n$C$_{60}$ with long-range orientational orders, justified by the grey circles in the top panels of Fig.\ 5(a). Moving to monolayer CsC$_{60}$ [Fig.\ 2(d)] and off-stoichiometric fullerides [Fig.\ 4], however, we see a dramatic change: the Fermi level deviates from $E_\textrm{i}$, moves upwards (right arrow) or downwards (left arrow), signifying $n$-type and $p$-type doping, respectively. Followed by the shift of $E_\textrm{F}$, a finite low-lying electronic states develops within the energy gap. This contradicts with stoichiometric Cs$_n$C$_{60}$, in which the $E_\textrm{F}$-centered flat bottom indicates vanishing electronic DOS there [Figs.\ 2(e), 2(f) and 3(e)-3(h)].

\subsection{Discussion}
\begin{figure}[h]
   \includegraphics[width=1\columnwidth]{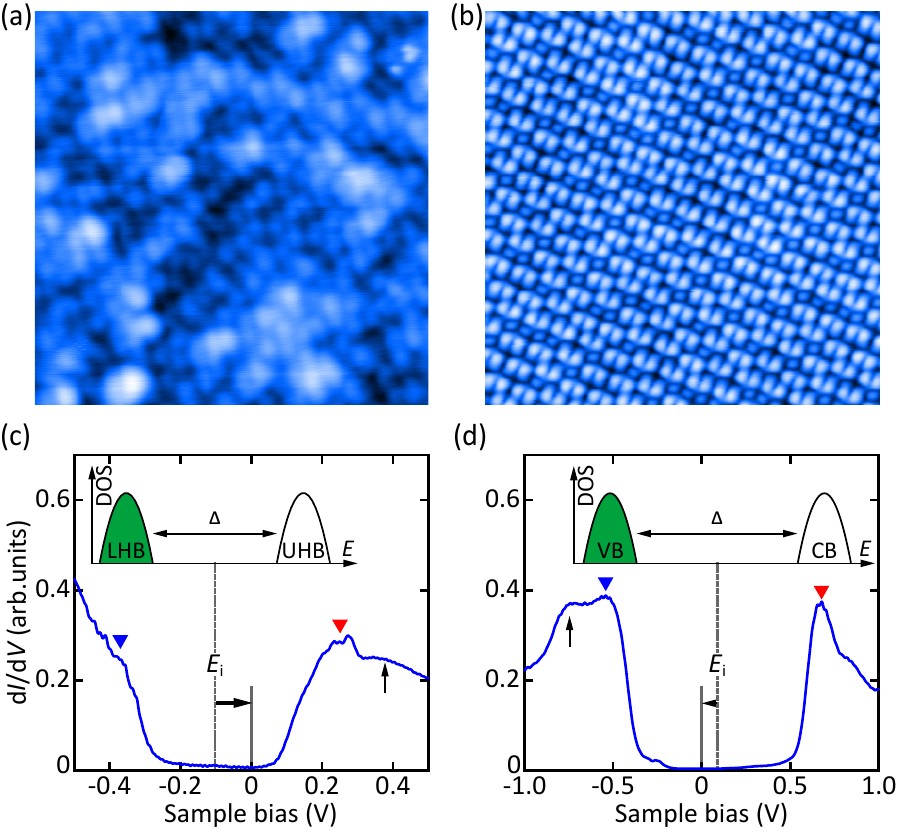}
\caption{Structure and electronic DOS in the off-stoichiometric Cs$_n$C$_{60}$ films. (a) Topography (20 nm $\times$ 20 nm, $V$ $=$ 1.0 V, $I$ $=$ 100 pA) and (c) conductance $dI/dV$ spectrum ($V$ $=$ 0.7 V, $I$ $=$ 100 pA) of monolayer Cs$_3$C$_{60}$ with excess Cs doping. The configuration of C$_{60}^{n-}$ is indiscernible due to the clustering of excess Cs. (b) Topography (15 nm $\times$ 15 nm, $V$ $=$ 1.0 V, $I$ $=$ 100 pA) and (d) $dI/dV$ spectrum ($V$ $=$ 1.0 V, $I$ $=$ 100 pA) of bilayer Cs$_4$C$_{60}$ with a deficiency $\delta \sim$ 0.06 of Cs. Here $\delta$ is estimated as the relative proportion of non-H:H orientated fullerenes in (b). The $E_\textrm{F}$ moves upwards (right arrow) and downwards (left arrow) relative to $E_\textrm{i}$ in (b) and (d), consistent with $n$- and $p$-type doping, respectively.}
\end{figure}

\begin{figure}[h]
   \includegraphics[width=1\columnwidth]{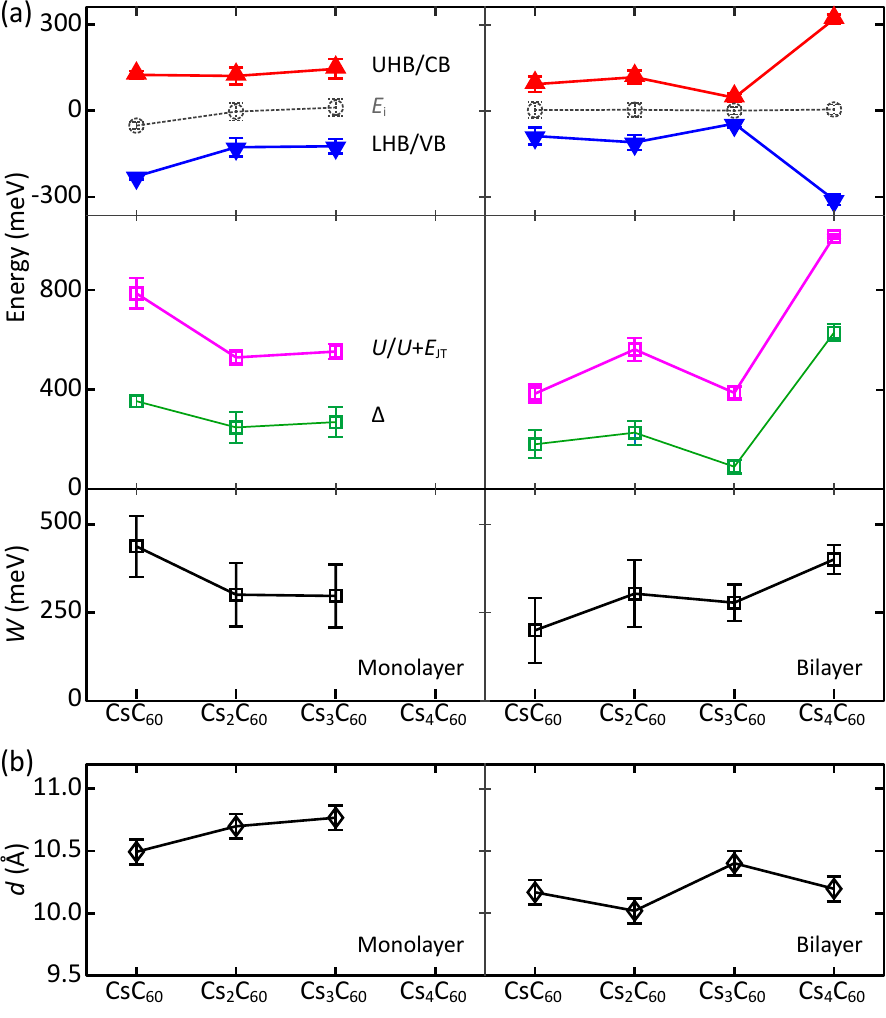}
\caption{Energy-band parameters and intermolecular spacing. (a) Onset energies of LHB/VB and UHB/CB (top panel), $U/U +$ $E_{\textrm{JT}}$, band gap $\Delta$ (middle panel) as well as bandwidth $W$ (bottom panel) in monolayer and bilayer Cs$_n$C$_{60}$ with varied $n$. Dashed circles denote the midgap energy $E_\textrm{i}$, half way between the UHB/CB and LHB/VB edges. (b) Dependence of intermolecular spacing $d$ on the doping level $n-$ and layer index. The errors indicate the standard deviations from multiple $dI/dV$ measurements in (a) or STM calibration uncertainty of $<$ 0.1 $\textrm{\AA}$ in (b).}
\end{figure}

Note that the third layer of C$_{60}$ behaves similarly with the bilayer one in the orientational ordering and electronic structure. In what follows, we discuss the mechanism for the insulating ground states in Cs$_n$C$_{60}$ fullerides with integer $n$. Upon doping with Cs in Cs$_n$C$_{60}$ (0$ < n <$ 6), the active $t_{1u}$ orbitals with triple degeneracy are partially filled and the simple band theory predicts a metallic character. In reality, however, the on-molecule Coulomb repulsion and JT instability, in spite of dynamic \cite{klupp2012dynamic,potovcnik2014jahn} or static \cite{wachowiak2005visualization,naghavi2016nanoscale}, are significant in stabilizing an insulating ground state against metallicity. The above observations of robust insulating gap around $E_\textrm{F}$, accompanied by the satellite peaks in electronic DOS (marked by black arrows) beyond the gaps, strongly hint at a possible lifting of the triply degenerate $t_{1u}$ frontier orbital via JT coupling in all C$_{60}^{n-}$ anions studied here (see inset of Fig.\ 1(e), top panel). Otherwise, the metal-like electronic DOS should have emerged in Cs$_n$C$_{60}$ fullerides with $n=$ 1, 2, 4. On the basis of JT coupling, it nowadays becomes straightforward to account for all the observed insulating ground states with the cooperative electronic correlations considered (see inset of Fig. 1(e), bottom panel). In Cs$_n$C$_{60}$ alkalides of even stoichiometry ($n=$ 2 and 4), the strong JT coupling splits the triply degenerate LUMO states into subbands and results into a charge-disproportionated insulator \cite{Intramolecular1997Ceulemans, iwahara2016orbital,Hoshino2017spontaneous,Ishigaki2018Spontaneously,Isidori2019charge}, whereas in odd stoichiometries of $n =$ 1 and 3 the strong electronic correlations further split one subband and lead to the Mott-insulating character. This has been well schematically illustrated by the inserted band structure in Figs. 2(d)-(f) and Figs. 3(e)-(h), with either lower/upper Hubbard bands (LHB/UHB) or valance/conductance bands (VB/CB) shown. Our results identify the universality of JT instability that cooperatively interplays with strong electronic correlations to stabilize the insulating states in all Cs$_n$C$_{60}$ studied ($n= 1-4$). The strong JT coupling overcomes the intramolecular Hund¡¯s rule electron repulsion and favorably stabilize low-spin magnetic states, matching excellently with the previous reports \cite{takabayashi2009disorder,ganin2010polymorphism,Fabrizio1997nomagneitic}. Moreover, our observations of sub-molecular orientational ordering support a static nature of the JT distortions at 4.8 K, at least in the millisecond time frame of STM imaging for single C$_{60}$ molecules \cite{Hands2010visualizaiton}.

Quantitatively, the energy gap $\Delta$ could be modelled as the difference between the Hubbard $U$ and the bandwidth $W$ in CsC$_{60}$ and Cs$_3$C$_{60}$, whereas in Cs$_2$C$_{60}$ and Cs$_4$C$_{60}$ the energy separation $E_{\textrm{JT}}$ between JT-split subbands becomes involved and it should be read as $\Delta$ $ = U + E_{\textrm{JT}}- W$ \cite{Lof1992band}. We estimate $U$ or $U + E_{\textrm{JT}}$ as the energy spacing between the two DOS peaks or kinks just outside the energy gap (marked by the triangles in Figs.\ 2(d)-(f) and Figs.\ 3(e)-(h)) and plot them as magenta squares in the middle panels of Fig.\ 5(a), from which the bandwidth $W$ could be readily extracted as well (bottom panel). As anticipated, the $W$ correlates inversely with the intermolecular spacing $d$ [Fig.\ 5(b)] and faintly fluctuates between 200 meV and 400 meV. By contrast, the $U$ or $U + E_{\textrm{JT}}$ alters significantly with the charge state $n-$ and layer index of C$_{60}^{n-}$ anions, being in phase with both $\Delta$ and $W$. This corroborates that the insulating energy gap $\Delta$ is controlled primarily by $U$ and $E_{\textrm{JT}}$, rather than $W$. As compared to bilayer Cs$_n$C$_{60}$ fullerides ($n = 1 - 3$), the monolayer exhibits a larger $U$ or $U + E_{JT}$, contradicting with K$_n$C$_{60}$ in Au substrate \cite{wang2008tuning}. One would easily envisage this discrepancy in view of the combined screening effects from nearby polarized C$_{60}^{n-}$ anions and the substrate. On Au(111), the substrate screening plays an essential role in Coulomb reduction, thereby leading to the smallest effective $U$ in the monolayer \cite{wang2008tuning}. However, the screening from graphene becomes negligibly small that only imposes a secondary effect on the Coulomb reduction. Alternatively, the increased nearest neighbors of C$_{60}^{n-}$ anions from 6 in the monolayer to 9 in the bilayer generate a more remarkable screening effect and thus smaller $U$, as observed. Furthermore, $U$ in Cs$_3$C$_{60}$ appears smaller than that in CsC$_{60}$ because the former has higher electron doping that more effectively reduce the Coulomb repulsion. In Cs$_2$C$_{60}$ and Cs$_4$C$_{60}$, both the band splitting and electronic correlations come into play, $U$ and $E_{JT}$ cannot be separately accessed in experiment. Nevertheless, the semi-quantitative analysis explicitly reveals that it is the combined JT coupling and electronic correlations to be responsible for the observed insulating ground states in Cs$_n$C$_{60}$ fullerides.

Finally, we comment on the electronic structure in off-stoichiometric fullerides. Starting with Cs$_3$C$_{60}$ Mott insulator and charge-disproportionated Cs$_4$C$_{60}$ insulator, one could easily understand the $n$-type and $p$-type behaviors in off-stoichiometric fullerides in Fig.\ 4, since an excess and deficiency of Cs dopants correspond to electron and hole doping, respectively. To fulfill the electronical charge neutrality, the $E_\textrm{F}$ is shifted upwards and downwards in monolayer Cs$_{3}$C$_{60}$ with excess Cs and bilayer Cs$_{3}$C$_{60}$ with Cs vacancies, resembling with the doping of semiconductors \cite{Rosenbaum1983metl}. Unlike atom-based semiconductors, however, the fullerene system exhibits additional rotational degree of freedom that is fragile to doping [cf. Fig.\ 3(d) and Fig.\ 4(b)]. Upon doping, the band structures, e.g., the $\Delta$ enhancement by 50 meV, might be modified as well. This accounts for the unexpectedly larger $U$ and $\Delta$ in monolayer CsC$_{60}$, considering that the absent long-range orientational ordering means a deviation of doping from unity [Figs.\ 2(a) and 2(d)]. Indeed, the in-gap states are observed in Fig.\ 2(d), resembling with the off-stoichiometry fullerides in Fig.\ 4. It is highly tempting to unravel the nature of these emergent low-lying electronic states, because they might prompt metallicity or even superconductivity in Cs-doped fullerides.

\section{Conclusions}

Our real-space imaging and spectroscopy of Cs-doped fullerides have established the unique correspondence among the doping level, molecular orientational ordering and electronic structure on the sub-molecular scale. The insulating ground states are invariably observed for integer filling and have their roots in the combined effects of JT coupling and electronic correlations. Additional electron and hole doping shift the $E_\textrm{F}$ upwards and downwards, accompanied by emergent in-gap states. The revealed similarity and distinction between the doping of fullerene and conventional semiconductors have general validity in other organic semiconductors and call for further experimental study to optimize superconductivity or other electronic properties.

\begin{acknowledgments}
The work is financially supported by the Ministry of Science and Technology of China (Grants No.\ 2017YFA0304600, No.\ 2016YFA0301004, No.\ 2018YFA0305603), the National Natural Science Foundation of China (Grants No.\ 11774192, No.\ 11427903, No.\ 11634007), and in part by Beijing Innovation Center for Future Chips, Tsinghua University.
\end{acknowledgments}

\begin{appendix}
\section{Phase separation and stoichiometry control}
\begin{figure}[b]
   \includegraphics[width=0.5\textwidth]{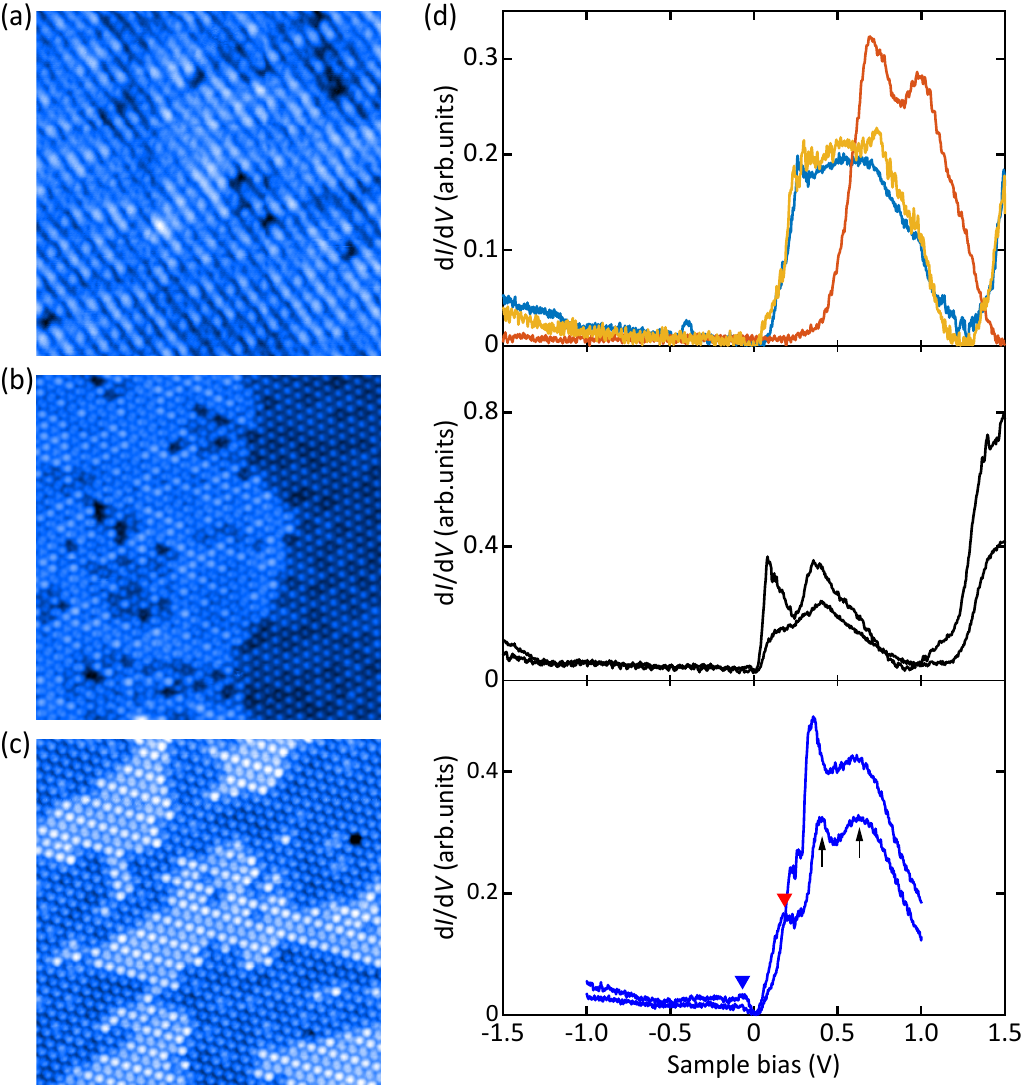}
\caption{Inhomogeneity and phase separation in fulleride films. (a)-(c) STM topographies ($V$ $=$ 1.0 V, $I$ $=$ 100 pA) of bilayer C$_{60}$ with the nominal Cs doping of 0.06, 0.4 and 0.9, respectively. Image size: (a) 20 nm $\times$ 20 nm, (b) and (c) 30 nm $\times$ 30 nm. At a sufficient doping of Cs, a phase separation occurs and is characterized by bright and dark regions. (d) Spatial-dependent tunneling $dI/dV$ spectra on Cs$_{0.06}$C$_{60}$ (top panel), the bright (middle panel) and dark region (bottom panel) of bilayer Cs$_{0.4}$C$_{60}$. Black and red arrows mark discrete DOS peaks caused by JT distortions and Coulomb interaction. The setpoint is stabilized at $V$ $=$ 3.0 V and $I$ $=$ 200 pA.}
\end{figure}

We start with the non-integer filled fullerides to track the doping dependence of STM topographies and differential conductance $dI/dV$ spectra. Figures 6(a)-(c) depict the representative STM images of bilayer C$_{60}$ with a nominal Cs doping of 0.06, 0.4 and 0.9, respectively. Evidently, the doping leads to great spatial inhomogeneity and phase separation. At $n\sim$ 0.06, a destruction of quasi 2 $\times$ 2 superstructure is apparent, followed by the emergence of spatial inhomogeneity in $dI/dV$ spectra (top panel of Fig.\ 6(d)). The LUMO electronic states at certain positions are shifted downwards (or equivalently an upward shifting of $E_\textrm{F}$), consistent with electron doping by Cs. As the doping is further increased, one can see that the STM topographies are divided into two distinct regions [Figs.\ 6(b) and 6(c))]. Both molecular and electronic structures in the bright regions are in analogy with those in Fig.\ 6(a), except for a larger upward shifting of $E_\textrm{F}$ (middle panel of Fig.\ 6(d)). On the other hand, the dark regions are characteristic of more ordered molecular orientations, and increase in area with Cs doping. The corresponding $dI/dV$ spectra present a fundamental difference from those of bright regions. A new energy gap is opened around $E_\textrm{F}$, as evident in the bottom panel of Fig.\ 2(d). Together with the discrete DOS peaks just above $E_\textrm{F}$ probably due to the JT splitting of $t_{1u}$ orbital (marked by the black arrows), we can comprehend the gap opening (separated by the red arrows) as a consequence of strong Coulomb interactions, as detailed in the main text. This hints at that the dark regions with sub-molecular orientational ordering correspond to a stoichiometric composition of CsC$_{60}$. Indeed, a sequence of molecular orientational orders are found and become dominant at the nominal Cs doping $n=$ 1 $\sim$ 4, as summarized in Figs. 2 and 3. This enables us to tune the stoichiometry of Cs$_n$C$_{60}$ fullerides and study their electronic structures.
\end{appendix}



%

\end{document}